%% file: main.tex
\documentclass[journal]{vgtc}                     % final (journal style)
\onlineid{0}

\vgtccategory{Research}

\title{Autark: A Serverless Toolkit for Prototyping\\Urban Visual Analytics Systems}

\author{%
  Lucas Alexandre, 
  \authororcid{João Rulff}{0000-0003-3341-7059},
  Talisson Souza, 
  \authororcid{Gustavo Moreira}{0000-0002-4762-7703}, 
  \authororcid{Daniel de Oliveira}{0000-0001-9346-7651},\\
  \authororcid{Claudio Silva}{0000-0003-2452-2295},
  \authororcid{Fabio Miranda}{0000-0001-8612-5805}, and
  \authororcid{Marcos Lage}{0000-0003-3868-8886}
}

\authorfooter{
  \item
  	Alexandre, Souza, de Oliveira, and Lage are with Universidade Federal Fluminense.
  	E-mail: \{danielcmo,mlage\}@ic.uff.br
  \item
  	Rulff and Silva are with New York University.
  	E-mail:  \{jlrulff,csilva\}@nyu.edu

  \item Moreira and Miranda are with the University of Illinois Chicago.
    E-mail: \{gmorei3,fabiom\}@uic.edu
}

\abstract{%
The development of visual analytics (VA) systems has traditionally been a labor-intensive process, balancing design methodologies with complex software engineering practices. 
In domain-specific fields like urban VA, this challenge is amplified by heterogeneous data streams and a reliance on complex, multi-service architectures that hinder fast development, deployment, and reproducibility.
Despite the richness of the urban VA literature, the field lacks a consolidated toolkit that encapsulates the core components of these systems, such as spatial data management, analytical processing, and visualization, into a unified, lightweight framework. 
In this paper, we introduce \autark, a serverless toolkit designed for the rapid prototyping of urban VA systems. \autark provides domain-aware abstractions through a self-contained architecture, enabling researchers to transition from design intention to deployed, shareable systems within hours.
Furthermore, \autark's structured, tightly scoped interfaces make it well-suited for AI-assisted coding workflows, where LLMs produce more reliable code when composing from well-defined abstractions rather than generating complex solutions from scratch. 
Our contributions are: (1) the \autark toolkit, a serverless architecture for rapid prototyping of urban VA; (2) a comparative study of LLM coding effectiveness with and without \autark; and (3) a series of usage scenarios demonstrating its capability to streamline the creation of robust, shareable urban VA prototypes.
\autark is available at \href{https://autarkjs.org/}{https://autarkjs.org/}
}


\teaser{
  \centering
  \includegraphics[width=\linewidth, alt={Recriation of urbane using Autark.}]{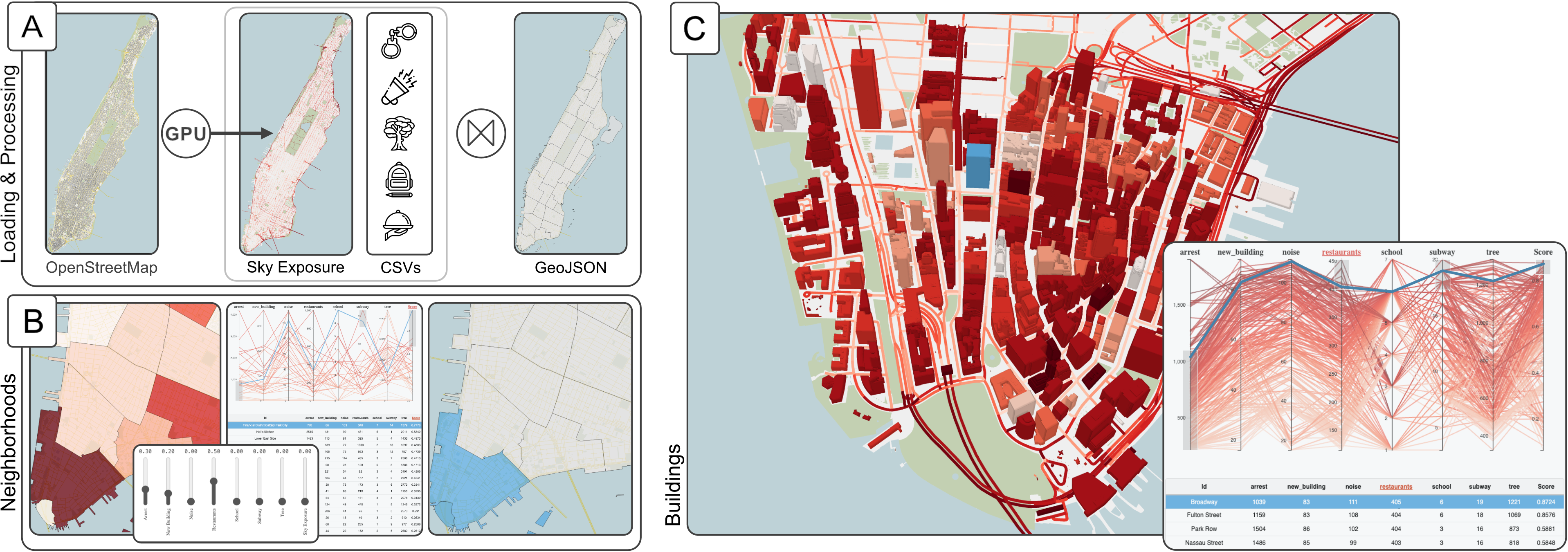}
  \caption{%
    Urbane~\cite{ferreira_urbane_2015} is an urban visual analytics system that supports multi-resolution exploration. It enables users to characterize urban areas based on data and identify sites for new development at the neighborhood and building levels. This Figure illustrates a remake of Urbane using \autark.
    \textbf{(A)} On initialization, the system fetches OpenStreetMap data and computes sky exposure values via the GPU. It also loads the Manhattan neighborhood GeoJSON and CSV files containing thematic data, such as arrests and noise complaints, and joins them to the neighborhood polygons.
    \textbf{(B)} Users can define weights for each attribute to compute a score in real-time using the GPU. A choropleth map, a parallel coordinates chart, and a table allow users to explore data distributions and perform interactive multi-attribute filtering. Selecting a neighborhood triggers a drill-down to the building level.
    \textbf{(C)} At the building level, the system joins all datasets to the building footprints within the selected neighborhood. The 3D map displays buildings color-coded by thematic data or scores. Users can identify sites of interest using the same linked and interactive visualizations available at the neighborhood level.
  }
  \label{fig:teaser}
}

\graphicspath{{figs/}{figures/}{pictures/}{images/}{./}} % where to search for the images

\usepackage{tabu}                      % only used for the table example
\usepackage{booktabs}                  % only used for the table example
\usepackage{lipsum}                    % used to generate placeholder text
\usepackage{mwe}                       % used to generate placeholder figures
\usepackage{ccicons}                   % package to be able to use icons from creative commons

\usepackage{mathptmx}                  % use matching math font
\usepackage{xcolor}
\usepackage{soul}
\usepackage{amsmath}
\usepackage[frozencache,cachedir=.]{minted}
\usepackage{tcolorbox}
\usepackage{amssymb}
\usepackage{pifont}
\usepackage[table]{xcolor}
\usepackage{makecell}
\usepackage{tabularx}
\usepackage{mathtools}

\definecolor{circlecolor}{HTML}{7570B3}
\definecolor{intelligencecolor}{HTML}{cb181d}
\definecolor{designcolor}{HTML}{238b45}
\definecolor{choicecolor}{HTML}{1f78b4}
\definecolor{codegreen}{HTML}{6a51a3}
\definecolor{myblue}{HTML}{54278f}

\tcbuselibrary{minted,skins}
\newtcblisting[auto counter]{autkcode}[1][javascript]{
  listing engine=minted,
  minted language=#1,
  minted options={fontsize=\small, breaklines},
  colback=white,
  colframe=white,
  frame hidden,
  enhanced,
  listing only,
  arc=0pt,
  boxrule=0pt,
  boxsep=0pt,
  left=0pt, right=0pt,
  top=0pt, bottom=0pt,
  toprule=0pt, bottomrule=0pt, leftrule=0pt, rightrule=0pt,
  before={\par\vspace{2pt}\noindent},
  after={\par\vspace{2pt}}
}

\newcommand{\autark}{Autark\xspace}
\newcommand{\myparagraph}[1]{\noindent\textbf{#1:}}
\newcommand{\hide}[1]{}
\newcommand{\highlight}[1]{\sethlcolor{yellow}\soul@hl{#1}}

\newcommand{\marcos}[1]{{\color{red}[Marcos: #1]}}
\newcommand{\daniel}[1]{{\color{purple}[Daniel: #1]}}
\newcommand{\joao}[1]{{\color{blue}[Joao: #1]}}
\newcommand{\fabio}[1]{{\color{orange}[Fabio: #1]}}

\begin{document}
%%%%%%%%%%%%%%%%%%%%%%%%%%%%%%%%%%%%%%%%%%%%%%%%%%%%%%%%%%%%%%%%
%%%%%%%%%%%%%%%%%%%%%% START OF THE PAPER %%%%%%%%%%%%%%%%%%%%%%
%%%%%%%%%%%%%%%%%%%%%%%%%%%%%%%%%%%%%%%%%%%%%%%%%%%%%%%%%%%%%%%%

\input{tex/01-intro}

\input{tex/02-related}
\input{tex/03-background}
\input{tex/05-autark}
\input{tex/06-examples}
\input{tex/07-benchmarks}
\input{tex/08-conclusion}
% \input{tex/xx-codes}
%
%% if specified like this the section will be omitted in review mode
% \acknowledgments{%
% }
%
\bibliographystyle{abbrv-doi-hyperref}
\bibliography{clean_refs2}

\end{document}

%% file: tex/01-intro.tex
\firstsection{Introduction} \label{sec:introduction}

\maketitle

The creation of visual analytics (VA) systems is at the cornerstone of visualization as a field.
Over the past decades, hundreds of systems have been proposed tackling domains, data, and problems as diverse as biological pathways~\cite{partl2013enroute} and transportation routing~\cite{guo2011tripvista}.
Given that one of the main objectives is a software artifact, this creation process is not only guided by design methodologies~\cite{munzner2009nested, sedlmair2012design}, but also heavily influenced by software engineering practices and tools.
This software engineering dimension has been shaped by the tools available to visualization researchers. Visualization libraries~\cite{bostock2011d3}, grammars~\cite{satyanarayan2016vega}, knowledge bases~\cite{moritz2018formalizing}, and collaborative canvases~\cite{observablehq} have each redefined what it means to \emph{prototype} a VA system: how fast an idea can be instantiated, validated with users, and discarded or refined.
Today, we are in the midst of what may be the most disruptive such shift yet: the rise of large language models (LLMs) and AI-assisted coding.
Tools and services have fundamentally altered the programmer's autonomy, compressing the distance between a design intention and a working implementation.
AI-assisted coding qualitatively changes what a single researcher can sustain alone; components that once required hours of careful engineering can now be generated and adapted much faster, shifting the bottleneck from \emph{writing code} to \emph{specifying intent}.
Yet, this new generative capability also surfaces new structural challenges, as LLMs produce more reliable code when composing from well-defined, reusable abstractions rather than generating complex solutions from scratch~\cite{stengel2024regal}.
A promising strategy is to then provide well-scoped software libraries that encapsulate complex low-level implementations into high-level abstractions that can be reused to build systems.
This approach carries a dual benefit. On one hand, consistent conventions and narrow interfaces reduce the search space an LLM must navigate, leading to more predictable and correct code generation. On the other hand, the same structure serves human developers: codebases built on coherent abstractions are easier to read, verify, and extend, a property that becomes critical in agentic workflows, where a developer must inspect and maintain code they did not write themselves.

The visualization community has long proposed general and reusable toolkits~\cite{bostock2011d3, heer2005prefuse}, frameworks~\cite{bostock2009protovis}, and component libraries~\cite{wongsuphasawat2020encodable} that encapsulate analytical, rendering, and interaction logic.
This approach of structured abstraction has, over time, lowered the barriers to visualization prototyping, with users assembling visualizations and systems from known primitives rather than writing from scratch, which is precisely the mode of development in which AI-assisted coding thrives~\cite{stengel2024regal}. 
These general libraries, however expressive, leave a gap for domain-specific tasks, as their data types, analytical pipelines, and visualization encodings fall outside any general approach.
Urban VA is a particularly salient case.
Cities generate heterogeneous data streams from trajectories, point-of-interest distributions, real-time sensor feeds, etc, each carrying its own spatial scale, temporal resolution, and analytical constraints~\cite{bonadia_visual_2023}.
Building VA systems on such data requires not only specific rendering capabilities but also a coherent set of domain abstractions: data loaders, spatial operators, and coordinated views tied to the urban data and the problem at hand.
Despite the richness of the urban VA literature~\cite{ferreira2024assessing} and several targeted contributions~\cite{miranda_state_2024}, the field lacks a consolidated, reusable toolkit for rapid prototyping, leaving researchers to reassemble the same foundational components with each new project, often leading to complex architectures where multiple disconnected components handle data ingestion, spatial aggregation, and queries, instead of lightweight, unified architectures that lower deployment overhead and improve reproducibility.
While effective at scale, this pattern has become the default, even for research prototypes where scalability might not be the primary concern.
The result is a growing body of systems that are difficult to reproduce, share, or maintain; prototypes that carry their own infrastructure dependencies, deployment steps, and operational overhead that rarely survive beyond the life of the project~\cite{ziegler2023need, freire2012computational}.

The aforementioned convergence of AI-assisted coding, library development, and simplified architectures opens a new design space for urban VA prototyping, enabling us to quickly move from intention to a deployed, shareable system within hours rather than weeks.
Realizing this potential, however, requires the necessary tooling that is simultaneously domain-aware, lightweight, and structured to leverage AI-assisted workflows.
To realize this opportunity, we present \autark, a serverless toolkit for rapid prototyping of urban visual analytics systems, designed around these three principles.
\autark integrates the core components of urban visual analytics systems into a high-level, structured, and unified codebase.
By consolidating essential components, such as spatial data management, parallel computing, and multi-scale visualization into a cohesive framework, \autark simplifies prototyping and experimentation. This integration allows developers and researchers to focus on defining analytical goals and interaction logic rather than managing complex multi-service infrastructures, while also providing developers and LLMs with a structured context environment to generate more reliable, higher-quality code to support development.
The contributions of this paper are threefold.
(1)~We first introduce the \autark toolkit, highlighting the advantages and limitations of its serverless architecture. (2)~We then present a set of usage scenarios implemented with \autark of systems that would traditionally be implemented using more complex architectures. (3)~Lastly, we provide a comparative evaluation of agentic system development with and without \autark, demonstrating the benefits of domain-specific abstractions for AI-assisted development.

%% file: tex/02-related.tex
\section{Related Work}

\myparagraph{Urban visual analytics systems}
Urban VA systems have been developed across a wide range of data domains, including human mobility~\cite{andrienko2017visual}, noise pollution~\cite{rulff2022urban}, environmental simulations~\cite{deng2019airvis}, public safety~\cite{garcia2021cripav}, and energy consumption~\cite{rodgers2011exploring}. 
Despite this diversity, these systems share the requirement for a non-trivial combination of sophisticated modules responsible for tasks such as data management, interactive visualization, and analytical processing.
TaxiVis~\cite{ferreira_visual_2013}, a foundational work on urban VA, exemplifies this complexity. To enable the interactive exploration of a massive dataset of taxi trips, the system requires a custom storage manager for efficient spatiotemporal retrieval, a visual query module, and multiple linked views, including maps with adaptive level-of-detail, heatmap overlays, scatterplots, and time series.
These systems become even more complex when three-dimensional data is required. Urbane~\cite{ferreira_urbane_2015}, for instance, was designed to support architects in evaluating the impact of new building developments in NYC. It integrates 2D and 3D data layers across components responsible for tasks such as impact analysis, landmark visibility, and sky exposure. 
Miranda et al.~\cite{miranda_state_2024} conducted a comprehensive survey of visual analytics for 3D urban data, covering applications including sunlight access, wind simulation, flood management, and view impact analysis. Their survey reveals that these systems must address challenges such as occlusion from dense geometry~\cite{elmqvist2008taxonomy}, navigation across spatial scales~\cite{chen2021urbanrama}, and integrating volumetric simulation outputs with the physical built environment~\cite{waser2014many}. These additional demands compound the integration burden, requiring yet more specialized modules and further increasing the complexity of system development.
Recent studies have introduced systematic methods for characterizing system complexity. Ferreira et al. \cite{ferreira2025va} developed VA-Blueprint, a methodology for extracting and organizing visual analytics components from research papers. Their analysis reveals that a typical urban VA system comprises 25 distinct components, and this complexity has been growing over time. These findings empirically demonstrate that building urban VA systems requires designing and integrating many specialized modules rather than just assembling standard parts.

\myparagraph{Tools for urban VA systems development}
Traditionally, GIS platforms, such as ArcGIS~\cite{booth2001getting} and QGIS~\cite{moyroud2018introduction}, have been the default tool for urban data analysis. 
Their monolithic design, however, hinders the development of new custom interactive systems, making them improper for the rapid prototyping cycles of visualization research~\cite{ziegler2023need}, where researchers need to experiment with different visualization designs, data processing techniques, and interaction mechanisms.
More recently, multiple web-based tools and libraries~\cite{maplibre, leaflet, openlayers, keplergl} have begun to address this limitation, offering modular, customizable building blocks for various components of a VA system. Deck.gl~\cite{deckgl} and Mapbox~\cite{mapbox} are geospatial visualization libraries providing layer-based spatial rendering capabilities. Moreover, general-purpose visualization grammars, such as Vega-lite~\cite{satyanarayan2016vega}, support geographic projections and mark types, thereby enabling the declarative specification of spatial visualizations.
Tailored to urban use cases, Urban Toolkit~\cite{moreira2023urban} proposes a grammar-based framework that integrates thematic and physical data layers across multiple spatial scales using a client-server architecture.
Interactive Urban VA also demands efficient data management techniques, particularly as datasets grow very large and span heterogeneous spatial and temporal scales.
Early systems such as imMens~\cite{liu2013immens}, Nanocubes~\cite{lins2013nanocubes}, and Falcon~\cite{moritz2019falcon} introduced precomputed data structures for scalable interaction with large spatiotemporal datasets.
More recently, Mosaic~\cite{heer2025mosaic} and its selection model~\cite{heer2025mosaicselections} provide an architecture that facilitates linking interactive views by constructing pre-aggregated data views, enabling low-latency cross-filtering.
The technology behind Mosaic is DuckDB~\cite{raasveldt2019duckdb}, which offers a WebAssembly variant that enables flexible deployment in web browsers and computational notebooks. 
In general, these data management tools are agnostic to urban-specific challenges and operations, such as heterogeneous geometries and physical urban layers, such as spatial joins between thematic data and building footprints. 
Autark addresses these gaps by consolidating spatial data management, GPU-accelerated computation, 3D map visualization, and abstract charts into a single serverless toolkit that runs entirely in the browser. 

\myparagraph{LLM in VA system development}
Enabling users to author VA system modules in natural language has long been a major goal in the visualization community.
Early efforts~\cite{gao2015datatone, sun2010articulate} relied on classical natural language processing techniques, such as rule-based methods and probabilistic grammars~\cite{setlur2016eviza}, while Flowsense~\cite{yu2019flowsense} employed semantic parsing to assist users in building visual dataflows. While successful in translating natural language into visual artifacts, these approaches were limited to predefined visualization types.
The emergence of LLMs has dramatically expanded what is achievable with natural language. Systems such as Urbanite~\cite{moreira2026urbanite}, Chat2VIS~\cite{maddigan2023chat2vis}, LIDA~\cite{dibia2023lida}, and ChartGPT~\cite{tian2024chartgpt} have demonstrated that LLMs are a promising avenue for automating visualization generation from free-form descriptions. This capability extends to more challenging scenarios: ChatVis~\cite{mallick2024chatvis} generates complete ParaView~\cite{ahrens2005paraview} scripts for 3D scientific visualization workflows, and NLI4VolVis~\cite{ai2025nli4volvis} enables real-time exploration and semantic editing of volumetric scenes using natural language.
More recently, researchers have begun leveraging LLMs not only to generate individual visual artifacts but also to orchestrate entire VA systems. LightVA~\cite{zhao2024lightva} decomposes high-level analytical goals into executable subtasks to facilitate coordinated visualizations, while Data-to-Dashboard~\cite{zhang2025data} automates end-to-end pipelines from raw data to interactive dashboards using modular agents that iterate and reflect on their decisions.
A fully general framework, however, capable of assembling VA systems across arbitrary domains may be impractical, as each domain brings its particularities. \autark fills the gap for a domain-specific tool that encapsulates most of the operations commonly needed to build urban VA systems, thereby not only supporting developers with a structured framework for defining their systems but also narrowing the design space the LLM must navigate to produce more robust and uniform VA systems.

%% file: tex/03-background.tex
\section{Design Principles}

\myparagraph{Lessons from urban VA development landscape}
Building VA systems for urban data presents multifaceted challenges arising from the intrinsic complexity of urban datasets, the architectural demands of such systems, the heterogeneous nature of their development ecosystem, and the high barriers to collaboration. All of these challenges shaped the design of \autark. 
Cities generate heterogeneous data streams, trajectories, sensor feeds, and environmental simulations, each carrying its own spatial scale, temporal resolution, and format, often reaching massive volumes~\cite{bonadia_visual_2023}. Despite the richness of the urban VA literature~\cite{miranda_state_2024, ferreira2024assessing}, the field lacks a consolidated toolkit for rapid prototyping, leaving researchers to reassemble the same foundational components with each new project. 
General-purpose visualization libraries~\cite{bostock2011d3, satyanarayan2016vega}, however expressive, leave a gap for domain-specific tasks whose data types, analytical pipelines, and visualization encodings fall outside any general approach. This gap motivates a design that must target the recurring data types, spatial scales, and analytical patterns that characterize urban VA \textbf{(Urban-specific)}, providing self-contained components that encode domain operations, such as spatial operations across multivariate data types, such as physical and thematic data, rather than requiring researchers to assemble these primitives from scratch.
From an architectural viewpoint, state-of-the-art urban visual analytics platforms typically integrate distinct modules for data ingestion and management, scalable computation and analytics, and both spatial and abstract visualizations~\cite{doraiswamy_spatio-temporal_2018, miranda_state_2024}. The orchestration of these modules may face different challenges depending on the adopted system architecture (e.g., client-server, microservices, etc.). For example, client-server architectures, widely used in web application development, must address challenges such as data transfer latency, state synchronization across distributed nodes, real-time performance during user interactions, and end-to-end security, particularly for sensitive urban datasets subject to privacy regulations.
These complex development infrastructures directly impact the practical results of urban VA systems. System deployment often becomes a difficult task that requires containerization and other specialized technologies, making tools hard to maintain and use. A toolkit designed to dissolve this overhead must propose a uniform architecture that facilitates distribution and reproducibility without requiring complex communication between heterogeneous services with different infrastructure requirements \textbf{(Web-first)}. \autark's serverless architecture is directly inspired by this observation. Its web-first stack, relying on WebAssembly and WebGPU, can be deployed as a static collection of modules, making systems inherently shareable and reproducible without environment configuration.
The design process of VA systems benefits from the ability to isolate and evaluate individual components independently, much like ablation studies. Replacing a spatial join strategy, swapping a rendering layer, or testing an alternative interaction model should not require restructuring the entire system. In practice, however, the tightly coupled architectures common in urban VA make such targeted experimentation difficult, discouraging exploration and creating barriers to collaboration with domain experts~\cite{moreira_curio_2025}, who cannot easily test ideas on isolated parts of the system. This motivates a design structured as a collection of loosely coupled, composable components with well-defined responsibilities and consistent interfaces \textbf{(Component-oriented)}. Researchers can adopt individual components without committing to the full toolkit, and swap them as requirements evolve without cascading changes.
 
\myparagraph{Lessons from agentic VA development}
Employing LLMs in the construction of urban VA systems has the potential to mitigate the challenges previously outlined, particularly by accelerating prototyping, bridging expertise gaps, and streamlining the use of heterogeneous development ecosystems. In fact, LLMs can efficiently generate boilerplate code, support API definition, improve code quality, and aid debugging, among other benefits. 
However, to fully exploit these benefits requires thorough context engineering, the deliberate curation of instructions that explicitly define the LLM's expected outputs, constraints, and success criteria. Calibrating the volume of information provided in the context is fundamental, as concise contexts yield outputs that fail to meet user goals, while large ones induce hallucinations due to diluted attention to key details.
In the context of urban visual analytics systems, one possible approach is to include general-purpose libraries useful for building these systems (e.g., spatial indexing libraries such as DuckDB, rendering frameworks such as Three.js, visualization libraries such as D3.js, or analytics tools such as scikit-learn) within the LLM. 
However, integrating several low-level libraries into the LLM context significantly increases token consumption and operational costs. This approach often results in a large, generic context that fails to capture domain-specific requirements or optimize code structures, ultimately reducing the LLM's effectiveness while increasing system maintenance costs.
Thus, the \autark's APIs were designed to be structured and narrow, reducing the surface area an LLM must reason over \textbf{(LLM-ready)}. Each component is documented with explicit contracts that can be included directly in a prompt context, benefiting both AI agents and human developers who must inspect and maintain code they did not write.
Reflecting on our experience designing urban VA systems~\cite{ferreira_visual_2013, ferreira_urbane_2015, miranda_shadow_2019, moreira_curio_2025, moreira2023urban, rulff2022urban} and surveying the field~\cite{ferreira2024assessing, miranda_state_2024}, we observe that the ambition of a design idea is rarely the bottleneck; the operational overhead of realizing it is. With \autark, our philosophy is that a toolkit should dissolve this overhead rather than manage it, allowing researchers to focus entirely on the design problem at hand.

%% file: tex/05-autark.tex
\begin{figure}
  \centering\includegraphics[width=1\linewidth]{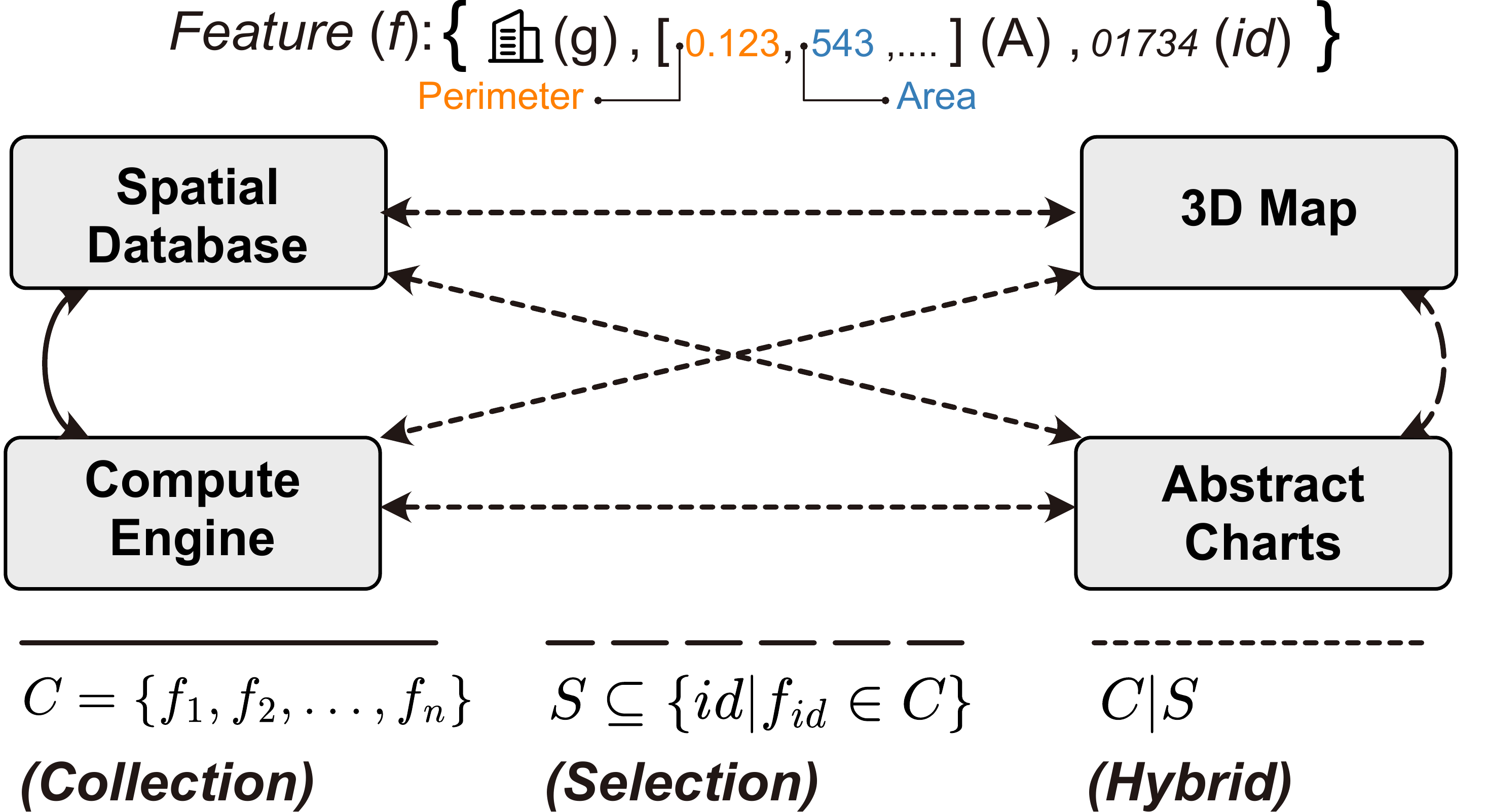}
  \caption{\autark's serverless architecture. The toolkit comprises four independent modules that can be freely combined via their APIs. It uses feature collections $C$ and selections $S$ in accordance with the feature-centric model. The image shows the data that flows through each connection. Hybrid connections exchange either collections or selections, depending on the direction of flow.}
  \label{fig:architecture}
\end{figure}

\section{The Autark Toolkit}
\label{sec:autark}

\autark is a toolkit designed to streamline the prototyping of urban VA systems through a serverless architecture that executes entirely in the web browser.
What makes its modules composable, both for human developers and for agentic workflows, is a programming model that governs how all components communicate.
We first describe this model, as it provides the conceptual foundation for understanding each module's design.
We then present \autark's components, illustrated in Figure~\ref{fig:architecture}: a spatial database that provides in-browser relational storage and querying of heterogeneous urban datasets built over DuckDB; a GPU compute engine built using WebGPU that executes analytical operations in parallel across geographic features; a 3D map view that renders vector and raster spatial data also using WebGPU; and an abstract charts library that provides visualizations built using D3.js.

\subsection{A unified model for data and interaction in urban VA}

Urban VA systems require two features to work in tandem: a coherent strategy for integrating heterogeneous data streams across spatial scales and formats, and a mechanism for coordinating interactions across multiple linked views.
Both problems in the urban setting can naturally be reduced to a common solution: geographical features describing the urban environment.
Urban data, regardless of its origin, can be expressed as a collection of features with associated properties, and user interactions, regardless of their visual form, can be expressed as selections over those same features.
\autark takes this observation as the foundation of its architecture model, making the geographical feature the atomic unit of the entire toolkit.

% , not just of data representation, but of computing, rendering, and interaction.

In most urban VA toolkits, data, rendering, and interaction each operate at different levels of abstraction: a database returns rows, a renderer consumes buffers, and an interaction produces a pixel coordinate or a bounding box.
\autark collapses these into a single level: everything operates on features.
In \autark, a spatial database ingests heterogeneous urban data and exports it as a feature collection; a compute engine executes analytical operations across features and writes results back as new feature properties; a map component renders each feature's geometry; and a chart component binds to feature properties. In this model, the developer's effort to convert, reshape, or transfer data between components is minimized; instead, they reason about features and feature collections, and the toolkit ensures that every component understands the data.

We define this design as the \textbf{feature-centric model for urban VA}.
This model's central claim is that the complexity of urban VA systems often arises from a single architectural mismatch: data and interaction operate at different levels of abstraction.
This mismatch is visible in the most influential VA programming models. 
In Vega-Lite, data is a flat array of tuples and interaction is expressed as a selection predicate that filters those tuples, a clean model for statistical visualization, but one that has no notion of geographic identity.
% : a neighborhood is a row, indistinguishable from any other row, and linking it across views requires explicit key joins.
%
In Mosaic, interaction is expressed as filtering predicates that translate user actions into SQL query constraints, but spatial selections must be expressed as a geometric predicate rather than as a reference to the features themselves.
deck.gl moves closer to a feature-centric model by organizing data into layers that can operate directly on features. However, this consistency does not extend across the full system: the base map in deck.gl is a tile layer that has no notion of features, so a building in a layer and the geometry underneath it are fundamentally two different kinds of objects.
In each case, the same translation layer reappears: between the data representation and the geographical entity, between the interaction event and the data it refers to, or both.

\autark avoids this by treating all data, including coastlines, streets, and buildings, as the same collection of features that drive computation and interaction.
There is no base map, no tile layer, and no boundary at which the feature abstraction breaks down: the same primitive flows from data ingestion to rendering to interaction.
In \autark, a feature $f=(g,A,id)$ is a triple consisting of a geometry $g$, a set of named thematic attributes $A=\{a_1, a_2, ..., a_n\}$ and a unique identifier $id$. For example, a building is a feature whose geometry $g$ is a polygon footprint, and whose attributes $A$ may include height and year that it was built.
A collection $C=\{f_1,f_2,...,f_n\}$ is a set of features that serves as the unit of exchange between all modules.
A selection $S\subseteq \{id|f_{id}\in C\}$ is a subset of feature identifiers drawn from a collection.
The model then supports two classes of operations, both defined over these primitives.
A data operation $\delta:C\rightarrow C$ takes a collection and returns a collection that may be filtered, enriched, aggregated or transformed, but keeping the representation of features as the unit of output.
%
% An interaction operation $i: C\times E \rightarrow S$ takes a collection and a user event $E$ (e.g., a click or brush) and returns a selection, i.e., the result of a user operation against the features in a collection and the identification of features within it.
%
An interaction operation $i : C \times E \rightarrow S$ takes a collection $C$ and a user event $E$ (e.g., a click or a brush gesture) and returns a selection, i.e., the subset of feature identifiers $S$ from $C$ targeted by the user.
Because both operations are defined over the same primitives, no translation is needed: a selection $S$ produced by an interaction in one view can be applied directly to any collection $C$ in any other view that shares the same identifier space.
% , without mapping, indirection, and knowledge of how the selection was produced.

This model offers three concrete advantages.
First, it is semantically natural: a selected neighborhood is always a neighborhood, not an index into a buffer, not a row identifier in a result set, not a bounding box in pixel space, and developers reason about the same objects at every stage of the pipeline, from querying to computing to rendering to interacting.
Second, it is composable: components can be freely combined because they share not just a data format but a common vocabulary for what the data means.
% , with no centralized broker required to orchestrate the exchange.
%
Third, it aligns naturally with how urban data is sourced in practice: the two most common sources of urban physical layer data, OpenStreetMap and GeoJSON, both organize their content as collections of geographical features. This uniformity extends to non-spatial sources: when tabular data is ingested, \autark can join it to existing spatial features, so that all modules operate on the same feature-centric representation regardless of the original format.
% , each carrying typed attributes.
%
Next, we describe how this model is operationalized in practice using \autark's components, each of which implements a subset of operations over a shared feature space.

\subsection{The Autark components}

% \autark is a toolkit that implements the previous model in TypeScript and that was designed to streamline the prototyping of urban visual analytics systems. 
%
\autark employs a serverless architecture, written in TypeScript, that executes entirely in the browser, eliminating backend dependencies and making deployment and distribution straightforward.
% for web applications.
%
As illustrated in Figure~\ref{fig:architecture}, \autark is composed of four core modules: a \textbf{spatial database} (Section~\ref{sec:autk-db}), a general-purpose GPU \textbf{compute engine} (Section~\ref{sec:autk-compute}), a \textbf{3D map} viewer (Section~\ref{sec:autk-map}), and a collection of \textbf{abstract charts} (Section~\ref{sec:autk-plot}). 
All modules are designed as standalone components and can be composed to handle complex usage scenarios. No centralized broker is required to orchestrate data exchanges between components. Components communicate via feature collections, the common communication protocol in \autark.
% and any component that understands a feature collection also understands the output of any other component.
% spanning from data ingestion to rendering.
% that uses the previously mentioned features for seamless integration 
%
% In \autark, there is no centralized broker orchestrating data exchanges between components. Components communicate via feature collections, and any component that understands a feature collection also understands the output of any other component.

% $\text{Spatial database}
% \xrightleftharpoons[C']{C}
% \text{Compute engine}
% \xrightleftharpoons[S]{C'}
% \text{Map \& charts}$

%%

\subsubsection{Spatial database} \label{sec:autk-db}

The spatial database module is the entry point of the feature-centric model. It provides in-browser relational storage and querying powered by DuckDB (compiled to WebAssembly), ingests heterogeneous urban datasets, and exports them as a collection of features $C$, the shared primitive that all downstream modules operate on.
In terms of the model's operations, the module is responsible for two tasks: constructing $C$ from raw urban data sources and implementing data operations $\delta: C \rightarrow C$ that filter, join, and aggregate features while preserving the features as the output units.

\myparagraph{Data injection} The spatial database module fetches and parses coastlines, parks, water bodies, streets, and buildings directly from OpenStreetMap. It relies on the OverpassAPI\footnote{\url{https://overpass-turbo.eu/}}, which supports retrieval of up to 10 million elements filtered by location, object type, tags, proximity, and other criteria. 
%
% Once fetched, each data layer is converted to GeoJSON and stored in DuckDB tables with appropriate spatial indexing.
Once fetched, each data layer is stored in an individual DuckDB table with appropriate spatial indexing.
After instantiating and initializing the database, to fetch OSM data, the user must define the base search area, the regions to load (e.g., neighborhoods), the output table name, the coordinate format, and the layers to extract. The following example loads all layers of Manhattan Island.
% , to use the OSM data loading API,
%
\begin{autkcode}[javascript]
const db = new AutkSpatialDb();
await db.init();

await db.loadOsmFromOverpassApi({
    queryArea: { 
        search: 'New York', 
        load: ['Manhattan Island'] 
    },
    outputTableName: 'table_osm',
    autoLoadLayers: {
        projection: 'EPSG:3395',
        layers: ['surface', 'parks', 'water', 'roads', 'buildings']
    }
});
\end{autkcode}
\noindent
The module can also load vector data (polylines and polygon sets) provided as GeoJSON, thematic urban data (point sets) in CSV, and raster data (imagery and heatmaps), accommodating both discrete data and continuous spatial fields in GeoTIFF.

\myparagraph{Spatial and OLAP queries}
Once ingested, the module supports data operations $\delta: C \rightarrow C$ that filter, enrich, and reshape the collection without ever breaking the feature as the unit of representation.
Linking urban datasets is essential in visual analytics for hypothesis testing, such as correlating incident reports with nearby street infrastructure. To support this, the spatial database module provides spatial joins (features within polygons) and nearest neighbor queries (features within a search radius), both with user-defined predicate functions for aggregation. 
\begin{autkcode}[javascript]
await db.spatialQuery({
    tableRootName: 'neighborhoods',
    tableJoinName: 'noise',
    spatialPredicate: 'JOIN', // 'JOIN', 'NEAREST'
    groupBy: {
        selectColumns: [
            {
                resultField: 'noise',
                column: 'noise_key',
                aggregateFn: 'count' // 'count', 'min', 'max', 'avg'
            }
        ]
    }
});
\end{autkcode}
\noindent
In exploratory tasks, the core data dimensions (i.e., spatial location, attribute values or event types, and timestamps) form a triad (where–what–when) where combining any two typically reveals patterns in the third~\cite{ferreira_visual_2013}. 
For this reason, \autark supports general queries for filtering by location (where), filtering and aggregating by attributes or categories (what), and slicing or grouping across time-varying datasets (when).

% In exploratory tasks, the core data dimensions (i.e., spatial location, attribute values or event types, and timestamps) form a triad (where–what–when) where combining any two typically reveals patterns in the third~\cite{ferreira_visual_2013}. 
%
% For this reason, \autark supports general SQL queries for filtering by location (where), filtering and aggregating by attributes or categories (what), and slicing or grouping across time-varying datasets (when). 
%
% These operations use DuckDB's built-in functionalities and its spatial extension to achieve low-latency results. 
%
% The possibilities and limitations of running a spatial database in the browser are discussed in Section~\ref{sec:results}.
%
%%%%%%%%%%%%%%%%%%%%%%%%%%%%%%%%%%%%%%%%%%%%%%%%%%%%%%%%%%%%%%%%%%%%%%%%%%%%%%%%%%%%%%%%%%%%%
%%%%%%%%%%%%%%%%%%%%%%%%%%%%%%%%%%%%%%%%%%%%%%%%%%%%%%%%%%%%%%%%%%%%%%%%%%%%%%%%%%%%%%%%%%%%%
\subsubsection{GPU compute engine} \label{sec:autk-compute}
%%%%%%%%%%%%%%%%%%%%%%%%%%%%%%%%%%%%%%%%%%%%%%%%%%%%%%%%%%%%%%%%%%%%%%%%%%%%%%%%%%%%%%%%%%%%%
%%%%%%%%%%%%%%%%%%%%%%%%%%%%%%%%%%%%%%%%%%%%%%%%%%%%%%%%%%%%%%%%%%%%%%%%%%%%%%%%%%%%%%%%%%%%%
%
The GPU compute engine also implements the data operation $\delta: C \rightarrow C$ in the feature-centric model. It takes a collection as input, executes analytical or render-based operations in parallel across its features on the GPU, and returns an enriched collection in which each feature carries new computed properties alongside its original attributes.
Urban visual analytics workflows typically require this kind of enrichment, combining physical layer features, such as street network segments, with derived thematic values, such as accumulated shadow durations or building visibility scores.
% , that are not present in the raw data and must be computed before visualization. 
The compute engine abstracts low-level WebGPU configuration details, providing a higher-level interface for data binding and shader definition, allowing users to focus on computation logic rather than GPU boilerplate.

\myparagraph{Compute programs}
The compute engine takes a feature collection $C$ as input. Each feature is mapped to a record in a data buffer consumed by a shader program, which is executed in parallel on the GPU. The metadata associated with each feature is explicitly bound to shader variables.
% in \autark's module API, making its attributes available within the shader implementation.
%
A full execution of the engine produces one (or a list of) output values per feature, which is then written back into the original structure as an additional attribute field, returning an enriched collection that preserves every feature and its original properties while extending each with the computed result.
%
% This output is immediately consumable by the map and chart modules without any format conversion, since it remains a collection of the same features the pipeline started with. 
The user supplies the analytical (from simple attribute transformations to complex domain-specific computations) or render-based logic (e.g., pixel counting for feature identification) as a WGSL code fragment that is injected into the shader, while the engine manages compilation, dispatch, and synchronization of GPU workloads. This design supports straightforward shader definition and rapid experimentation with analytical models 
The following example illustrates a simple computation that calculates the volume of each building and writes the result as a new feature property.
\begin{autkcode}[javascript]
const cp  = new AutkComputeEngine();
const enriched = await cp.analytical({
  collection, // input feature collection
  variableMapping: {
    x: 'area',
    y: 'height',
  },
  resultField: 'volume',
  wgslBody: 'return x * y;'
});
\end{autkcode}
%
% In summary, based on the defined inputs (attribute bindings) and outputs (new fields to be added to the features), the user is allowed to write a piece of code that is injected into the WGSL shader executed by the GPU, while the engine manages compilation, dispatch, and synchronization of GPU workloads.

\subsubsection{3D map visualization} \label{sec:autk-map}
The 3D map module operationalizes two roles in the feature-centric model. As a renderer, it consumes a collection $C$ and maps each feature's geometry and attributes to visual encodings on screen. 
As an interaction surface, it implements the operation $i: C\times E \rightarrow S$: user events trigger a picking operation that identifies the targeted features and emits a selection $S$ of their identifiers, which any other module can consume to drive coordinated updates. 
Unlike traditional tiled mapping libraries, which treat the map as a collection of pre-rendered layers without providing access to the underlying data, \autark renders all geometry (e.g., coastlines, streets, buildings, and thematic overlays) from the same feature collections that drive computation and interaction, so the feature abstraction is never broken at the rendering boundary.
The map module accepts data in GeoJSON and GeoTIFF formats, loaded either from files or from the spatial database module. 
\begin{autkcode}[javascript]
const collection = ...; // input feature collection

const map = new AutkMap(canvas);
await map.init();

map.loadCollection('neighborhoods', collection);
map.draw();
\end{autkcode}
\noindent
Once loaded, all data (point sets, polylines, polygons, and raster data) are represented as triangle meshes, ensuring a consistent, fast rendering pipeline regardless of the underlying geometry.
The viewer renders layers sequentially according to their loading order, enabling a multi-layered composition of heterogeneous data. 
%
% An analyst can overlay OpenStreetMap coastline and street network data to create mobility-focused visualizations, add 3D building data to study urban density and zoning patterns, or incorporate contextual layers such as neighborhood boundaries or park locations to support more targeted spatial analyses.

\myparagraph{Data encoding and map styling}
The map visualization module enables mapping feature attributes or aggregated thematic data (e.g. generated through the spatial database module), directly to the layer's color channel. This supports the creation of visualizations such as heatmaps and choropleth maps, and enables visual representation of dynamic highlighting via brushing operations across coordinated views.
\begin{autkcode}[javascript]
const collection = ...; // input feature collection

map.updateThematic('neighborhoods', collection, (f: Feature) => {
    return f.properties.sjoin.count.noise;
};);
\end{autkcode}
\noindent
The module also allows users to define custom themes by specifying color styles for each layer, supporting both neutral palettes (e.g., for background context) and saturated schemes (e.g., for emphasizing layers of interest).
% The module also allows users to define custom themes by specifying color styles for each layer. This flexibility is essential because, depending on the application, users must choose between highly saturated or neutral palettes. Neutral tones, such as shades of gray, are ideal when the map serves as a background canvas for other visualizations. Conversely, saturated colors are an effective design choice to highlight specific pieces of information that require attention during analysis.

\begin{figure}
  \centering\includegraphics[width=\linewidth]{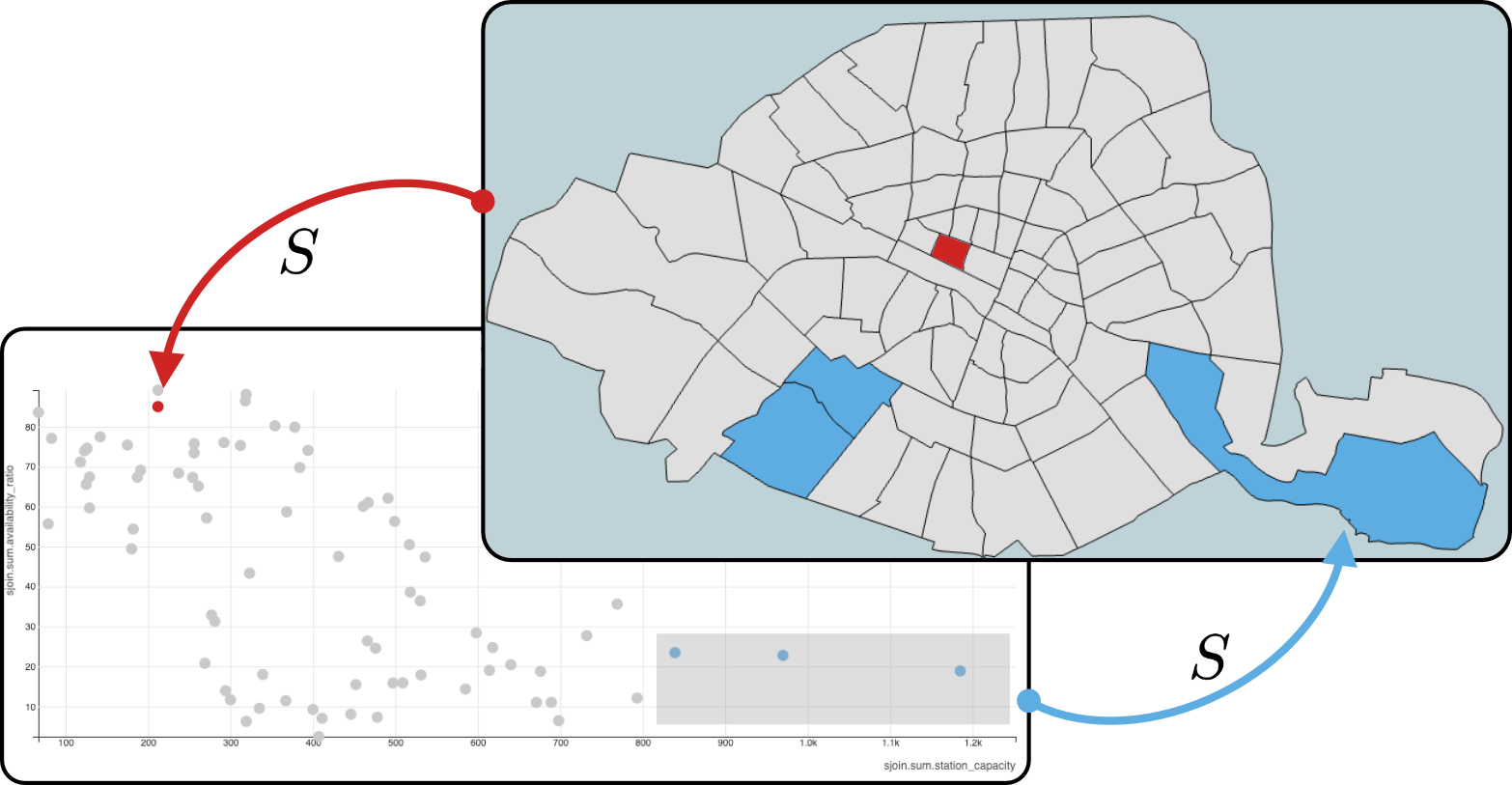}
  \caption{\autark's map and abstract charts modules facilitate the implementation of linked visualizations. Features selected on the map or brushed in the charts emit events containing selections ($S$) that other visualizations can react to. In this example, a map of Paris neighborhoods is joined with a city-wide bike station dataset. A scatter plot displaying station capacity versus availability ratio allows users to identify neighborhoods with high capacity but low availability, which are highlighted on the map. Conversely, selecting a neighborhood on the map immediately highlights its corresponding points in the scatter plot.}
  \label{fig:interaction}
\end{figure}

\myparagraph{Feature-Level interactivity}
Any layer can be configured to listen to interactions, triggering a picking operation that identifies the selected features by their identifiers. 
The module then fires an event $E$ carrying the resulting selection $S$, which other modules can capture through an event-driven architecture. In the example shown in Figure~\ref{fig:interaction}, selecting a neighborhood on the map (Palais Royal, Paris, in red) immediately highlights its corresponding data point in the scatter plot. This bidirectional flow ensures that selections made on the map trigger real-time updates in abstract charts, while interactions within those charts can, in turn, signal the map to highlight specific geographic features. 
% Because $S$ is expressed as a set of feature identifiers drawn from the shared collection, no translation is needed at either end of this exchange; the map and the chart speak the same language by construction.
%
% // picking activation for the 'neighborhoods' layer
% map.updateRenderInfo('neighborhoods', 'isPicking', true);

\begin{autkcode}[javascript]
// picking event callback definition
map.events.addListener(MapEvent.PICKING, (selection: number[]) => {
    // event handler
});
\end{autkcode}

%%%%%%%%%%%%%%%%%%%%%%%%%%%%%%%%%%%%%%%%%%%%%%%%%%%%%%%%%%%%%%%%%%%%%%%%%%%%%%%%%%%%%%%%%%%%%
%%%%%%%%%%%%%%%%%%%%%%%%%%%%%%%%%%%%%%%%%%%%%%%%%%%%%%%%%%%%%%%%%%%%%%%%%%%%%%%%%%%%%%%%%%%%%
\subsubsection{Abstract charts} \label{sec:autk-plot}

The abstract charts module shares the same two roles as the 3D map in the feature-centric model: it consumes a collection $C$, binding directly to feature attributes for rendering, and implements the interaction operation $i: C\times E\rightarrow S$, translating user events into selections $S$ of feature identifiers that any other module can consume.
By operating directly on features and accessing feature attributes from the properties object, the module eliminates the need for data format transformation and keeps a single source for both spatial and abstract visualizations. 
%
% The same collection that drives the map also drives the chart, and a selection produced in either is immediately legible to the other.
%
\autark provides multiple chart types (bar charts, scatterplots, line charts, parallel coordinates, etc.), selected based on the most commonly used chart types in the Urban VA literature.
% standard chart types built upon D3.js, including bar charts, scatterplots, line charts, and parallel coordinates.
%
\begin{autkcode}
const plot = new AutkChart(plotDiv, {
    type: 'barchart'
    collection, // input feature collection 
    labels: { axis: ['name', 'area'], title: 'Neighborhoods area' },
    events: [PlotEvent.CLICK]
});    
\end{autkcode}

\myparagraph{Extensibility} 
The library enables custom visualizations through a single draw method, within which the developer defines the mapping between feature attributes and visual elements using standard D3.js code and data join patterns. This approach offers two key advantages. First, updates are handled natively. When the underlying collection changes, for instance, after a new spatial join or a GPU computation, the library automatically synchronizes the updated attributes with the existing SVG elements, since it operates on the same feature collection rather than a transformed copy. Second, developers can reuse most of their existing D3.js implementations with minimal integration overhead, as the only requirement is that visual marks store the feature(s) identifier(s) they represent.

\myparagraph{User interactions} 
%
% The charts library handles user interactions through two CSS classes: \textit{.autkbrush}, which defines chart areas where interactions are captured, and \textit{.autkmark}, which labels individual SVG elements available for selection. 
%
When instantiating a chart, users can toggle between click and brush interaction modes. In click mode, the selected mark is the element that the user directly clicks. In brush mode, they are the elements in the brushed area. 
% In brush mode, D3.js brush objects are configured within \textit{.autkbrush} areas to capture all \textit{.autkmark} elements falling within the selection boundaries. 
In both cases, the module extracts the selection $S$ and broadcasts it using the same event-driven strategy used by the 3D map module. 
\begin{autkcode}[javascript]
// click event callback definition
plot.events.addListener(PlotEvent.CLICK, (selection: number[]) => {
    // event handler
});
\end{autkcode}

\noindent 
The previous example shows how to register a callback that receives the identifiers of the selected features when a click event occurs. Because $S$ is a set of feature identifiers drawn from the shared collection, it can be applied directly to the map, or to any other module, without translation. 
%
%
% \noindent
% As illustrated in Figure~\ref{fig:interaction}, selections made on abstract charts trigger real-time updates on the map, which highlights the corresponding geographic features, ensuring that statistical patterns discovered in the charts can be localized and examined within their spatial context.

% \fabio{Data: Diversity of data hence GeoJSON as a common data format}

% \fabio{Interaction: Event-driven calls to support user interaction}

% \fabio{Components: Modularity}

% \fabio{Components at the base; events between them to support user interactions; data to transfer information}

%% file: tex/06-examples.tex
\begin{figure*}[t]
  \centering\includegraphics[width=\linewidth]{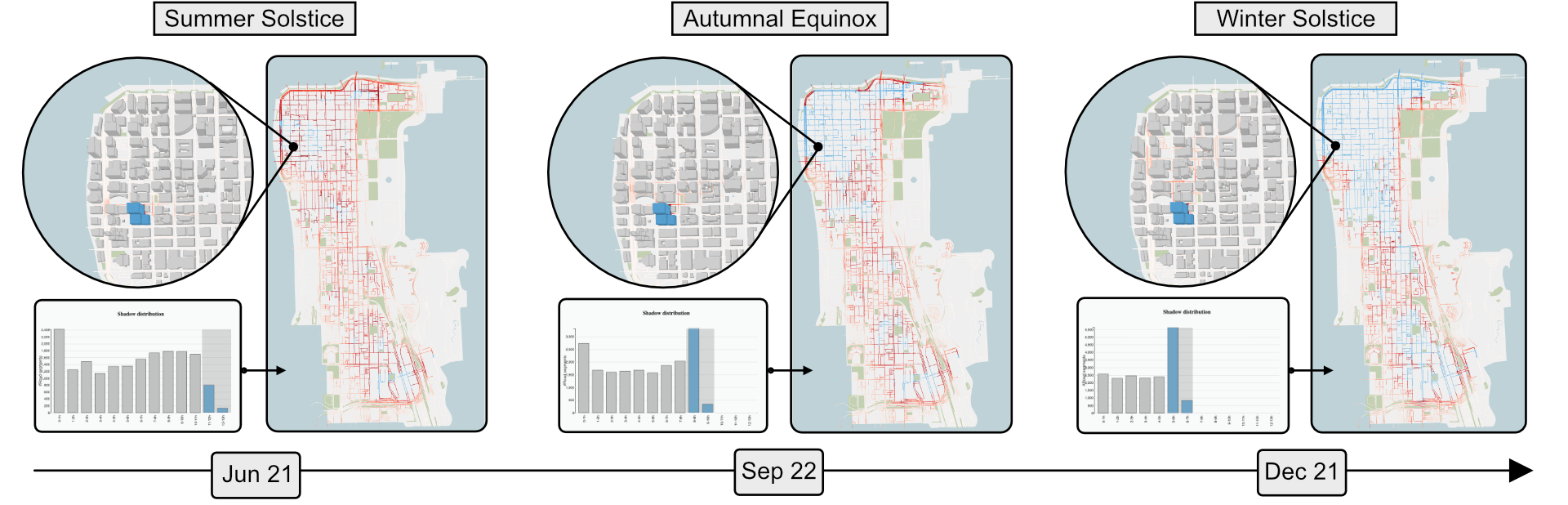}
  \caption{Shadow accumulation in the Chicago Loop during the Summer Solstice (June 21), Autumnal Equinox (September 22), and Winter Solstice (December 21). Developed with \autark the built interactive system integrates linked histograms and a map view with a database of precomputed city-wide shadow data. The histogram shows the distribution of road segments by shadow duration; users can brush specific time ranges on the chart to highlight the corresponding segments in blue on the map. These spatial views show that the number of roads remaining in shadow for extended periods increases as the season progresses toward winter. To provide granular insights, the system can compute shadows cast by a user-selected building onto nearby streets. These local values are calculated in real time using a ray-casting algorithm executed by the GPU compute engine.}
  \label{fig:shadows}
\end{figure*}
\section{Evaluation} \label{sec:examples}

We evaluate Autark along three dimensions. First, we present a series of usage scenarios demonstrating that Autark can achieve functionality comparable to urban VA systems traditionally built on complex, multi-service infrastructures. Second, we report performance benchmarks measuring data loading and spatial join scalability to assess the viability of a browser-based architecture for urban VA. Third, we conduct a controlled experiment comparing agent-based implementations of urban VA tasks with and without Autark to evaluate the toolkit's impact on agentic development workflows.

\subsection{Usage examples}
\label{subsec:usage_examples}

\myparagraph{Serverless remake of Urbane~\cite{ferreira_urbane_2015}}
Figure~\ref{fig:teaser} shows a remake of Urbane built with \autark. Urbane is a multi-resolution (neighborhood and building levels) VA system designed to support data-driven decision-making in urban development, enabling architects and planners to explore neighborhood characteristics across multiple integrated thematic datasets.
% selecting initially a neighborhood and then a building to plan a new development.
%
The original system followed a client-server architecture, relying on preprocessed OSM data for the physical layers and precomputed spatial aggregations to associate thematic data with neighborhoods and buildings.
In this reimplementation, all data is dynamically fetched and handled by the system using the spatial database module. OSM data containing the physical layers, a GeoJSON of Manhattan's neighborhoods, and CSV files of the thematic data of interest are fetched by the toolkit when the system is initialized (Figure~\ref{fig:teaser}(A)).
Once all data is loaded, we run a render-based program using the GPU compute module to estimate sky exposure, i.e., the percentage of the sky visible along the streets. Then, \autark performs spatial joins between the OSM buildings layer and the neighborhoods layer to determine which buildings belong to each region. Then, thematic and dynamically computed datasets are joined with the neighborhood layer so that each polygon is associated with its number of arrests, new buildings, trees, noise complaints, restaurants, schools, subway entrances, and the average sky exposure within its area.
The loaded data can be initially explored at the neighborhood level using three visualizations (Figure~\ref{fig:teaser}(B)): a choropleth map, a parallel coordinates chart, and a table, built using \autark's 3D-Map and abstract charts modules. All visualizations are linked.
% , and interacting with one view affects the others.
%
Once the user identifies a neighborhood of interest, the system allows them to drill down to the building level and explore the best location for their development.
% within the area. 
When the drill-down operation is triggered, the system first queries \autark's database to select buildings within the chosen area, then joins all datasets with the building footprints using a nearest-neighbor spatial operation.
The selected buildings and all their associated data can be visualized on the map using the color channel for buildings, as well as in the parallel coordinates and table visualizations previously used at the neighborhood level (Figure~\ref{fig:teaser}(C)). When interacting with the charts, the user can highlight the buildings that best fit their selection criteria.
Since \autark also allows running code in parallel for each feature of a given layer using the General-purpose GPU computation module, we also implemented in this Urbane remake a dynamic feature score based on user-defined weights through the visual interface (Figure~\ref{fig:teaser}(B)). Once the user defines the weights for each attribute of interest, the system instantiates the GPU computation module and recalculates the score for each element in real time.
The score computation is available at both the neighborhood and building levels.
Although the system remake does not faithfully reproduce the original system, it implements all core functionalities. If compared to the original implementation, which comprised thousands of lines of code, a complex architecture, and a heterogeneous development ecosystem spanning multiple programming languages and dependencies, the \autark-based implementation achieves the same functionality in under 350 lines of code, runs entirely in the browser, relies solely on TypeScript, and is straightforward to deploy and extend. A live version of this example can be found online at \href{https://autarkjs.org/usecases/live/urbane.html}{https://autarkjs.org/usecases/live/urbane.html}.

\begin{figure*}[t]
  \centering\includegraphics[width=\linewidth]{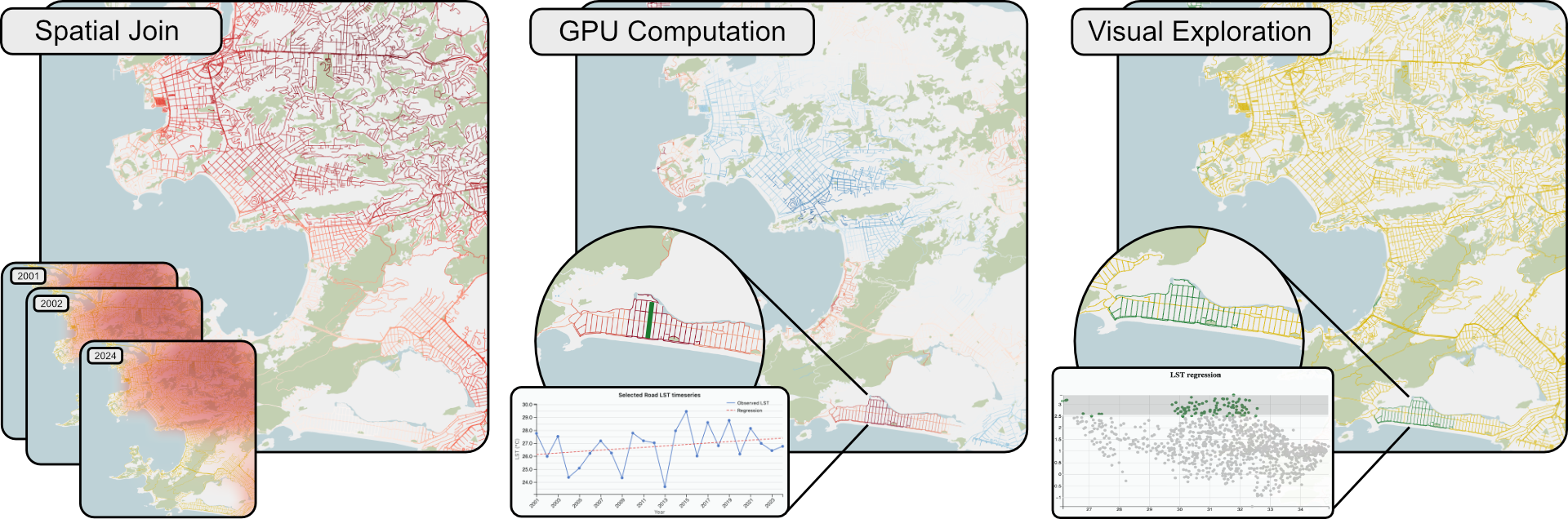}
  \caption{Analysis of urban heat trends in Niterói (2001–2024) using \autark. The system processes NASA LST raster data to enrich road segments with average summer temperature over the years (left column). \autark's compute engine, in parallel, calculates linear regressions for each road segment to map temperature trend (slope) and base heat (intercept) across the city. Line charts showing the heat data and the regression lines for picked roads can be produced dynamically (middle column). Finally, a scatter plot showing the slope and intercept values of all roads in the city is constructed to enable global exploration of heat areas. The chart's brushing interaction allows users to highlight areas of significant warming on the map, such as the Piratininga neighborhood, where recent urban expansion correlates with rising temperatures (right column).}
  \label{fig:heat}
\end{figure*}
\myparagraph{Accumulated shadows in Chicago}
Accumulated gross shadow is a measure that quantifies the total time a specific location is in shade during a given time interval~\cite{miranda_shadow_2019}. By capturing the cumulative duration of shadows across regions, this metric helps planners evaluate the impact of building developments on the environmental quality of public spaces, informing decisions about building height regulations and the preservation of sunlight access.
% , a metric that helps city planners evaluate the impact of building developments on the environmental quality of public spaces.
% This metric is important for urban design because shadows significantly influence the environmental quality of public spaces such as parks and streets. While shade can be beneficial for pedestrian comfort during hot summer months, it can also be detrimental by inhibiting vegetation growth or reducing the potential for solar energy. By computing gross shadow, city planners and architects can evaluate the impact of new building developments to ensure a balance between urban density and the public right to light.
%
Using \autark, we developed an application 
% (with ~300 lines of code)
% to analyze shadow patterns in the Chicago Loop during three seasonal markers: the summer solstice (June 21), the autumnal equinox (September 22), and the winter solstice (December 21), as shown in Figure~\ref{fig:shadows}. The workflow began by loading OpenStreetMap data layers for roads and buildings, along with the accumulated shadows dataset\footnote{\url{https://osf.io/4tqp9/overview}}. We then leveraged \autark's spatial database module to join the shadow values with the street network data. Finally, we used the \autark abstract charts module to generate a histogram of the number of street segments in shadow for durations ranging from 1 to several hours per day, facilitating a seasonal comparison of urban solar access.
%
to analyze shadow patterns in the Chicago Loop during three seasonal markers (Figure~\ref{fig:shadows}): the summer solstice (June 21), the autumnal equinox (September 22), and the winter solstice (December 21). The workflow began by loading OpenStreetMap data layers for roads and buildings, along with the accumulated shadows dataset\footnote{\url{https://osf.io/4tqp9/overview}}, which encodes the total daily duration of solar occlusion for each street segment across the three dates of interest. We then leveraged \autark's spatial database module to join the shadow values with the street network data, associating each road segment with its corresponding shadow duration. Finally, we used the \autark abstract charts module to generate a histogram of the number of street segments in shadow for durations ranging from 1 to several hours per day, facilitating a seasonal comparison of urban solar access.
%
% The system utilizes \autark's support for linked visualizations to implement a brush interaction on the histogram. As illustrated in Figure~\ref{fig:shadows}, an analyst can select specific time ranges to immediately identify and highlight corresponding road segments in blue on the map. These coordinated views reveal a clear seasonal trend: the number of roads experiencing prolonged solar occlusion increases significantly as the year progresses toward the winter solstice.
%
The system utilizes \autark's support for linked visualizations to implement a brush interaction on the histogram. As illustrated in Figure~\ref{fig:shadows}, an analyst can select specific time ranges to immediately identify and highlight corresponding road segments in blue on the map, enabling direct spatial inspection of the selected distribution. These coordinated views reveal a clear seasonal trend: the number of roads experiencing prolonged solar occlusion increases significantly as the year progresses toward the winter solstice.
%
% To take the analysis further, the application leverages the 3D map picking interaction and the \autark compute module to allow users to select a specific building and calculate its individual contribution to the accumulated shadows on surrounding streets. This feature, shown in the circular zoom insets in Figure~\ref{fig:shadows}, employs a ray-casting approach at an \textit{hourly resolution} for the day of interest. More precisely, for each 20-meter road segment, the system casts a ray toward the sun at every hour; if the ray intersects the selected building, it is recorded as a potential shadow source for that segment at that time. These local values are calculated in real time. 
%
To take the analysis further, the application leverages the 3D map picking interaction and the \autark compute module to allow users to select a specific building and calculate its individual contribution to the accumulated shadows on surrounding streets. This feature, shown in the circular zoom insets in Figure~\ref{fig:shadows}, employs a ray-casting approach at an \textit{hourly resolution} for the day of interest. More precisely, for each 20-meter road segment, the system casts a ray toward the sun at every hour; if the ray intersects the selected building, it is recorded as a potential shadow source for that segment at that time. These local values are calculated in real time, allowing analysts to interactively probe the shadow contribution of individual structures.
We observe that this computation result represents a lower bound, as other structures could be positioned between the building of interest and the segment. We argue that this approach ensures the estimated contribution is never lower than reality. This example can be found online at \href{https://autarkjs.org/usecases/live/shadows.html}{https://autarkjs.org/usecases/live/shadows.html}.

\myparagraph{Heat islands in Niterói}
This use case demonstrates \autark's ability to process raster geospatial datasets. To do so, we developed an application to monitor trends in urban land temperatures. The analysis uses NASA Land Surface Temperature (LST) data covering the period from 2001 to 2024 for the city of Niterói\footnote{\url{https://shorturl.at/R0G5v}}. We preprocessed the original raster files to create a multi-band TIFF, with each band storing the average summer temperature (i.e., data from December, January, and February) for a specific year, resulting in a compact representation of the dataset's full temporal extent. As illustrated in the first column of 
Figure~\ref{fig:heat}, we first use the spatial database module of \autark to load the multi-band TIFF file and perform a spatial join between the road network and the temperature time series, enriching each street segment with a time series of summer heat values that captures its thermal history over the observation period.
Next, to identify areas with increasing surface temperatures, we used \autark's compute engine to calculate linear regressions of surface temperature values for each road segment in parallel. The regression is computed independently for each segment, taking full advantage of the GPU's parallelism to process the entire road network simultaneously. The slope of the regression indicates the trend in temperature variation, while the intercept encodes the base temperature values and may help locate the hottest areas. In the middle column of Figure~\ref{fig:heat}, we see the road network of Niterói colored by the regression slope (using a diverging scale, from blue to red). Negative slope values mean that temperatures in the region decreased from 2001 to 2024. Positive values mean that the temperature is increasing in the region.
We also used the feature-picking functionality of \autark's 3D-map module to implement a dynamically updated line chart that displays the temperature time series and the fitted regression line for segments of interest, enabling detailed investigation of localized thermal trends and specific anomalous situations.
Finally, we created a scatterplot to facilitate a global exploration of heat areas over the city. The chart implements a brush interaction that allows users to highlight roads on the map based on the slope and intercept values of interest. By jointly analyzing the slope and intercept of each road segment, users can distinguish between areas that are both hot and warming 
from those that are warming but starting from a lower baseline. Interactively exploring the visualizations, we identified that one of the regions with the highest increases in temperature was the Piratininga neighborhood (see the last column of Figure~\ref{fig:heat}), where intense urban expansion and the construction of several new buildings have occurred in recent years. A live version of this example is available at \href{https://autarkjs.org/usecases/live/heat.html}{https://autarkjs.org/usecases/live/heat.html}.

%% file: tex/07-benchmarks.tex
% \section{Results} \label{sec:results}

% \fabio{TODO: Needs an intro paragraph saying that there are two evaluations: benchmarks and the agentic AI stuff. Maybe we reduce the benchmark part to just one paragraph? And without the plot?}
\newpage
\subsection{Performance}
\label{subsec:performance}
%
% Three benchmark experiments (performed in an M1 Mac Studio, with 48GB of RAM) were conducted to evaluate the feasibility of \autark's serverless approach.
% %
% In the first, the system's loading time was assessed by incrementally loading the Manhattan neighborhood's data, ranging from 1 to 10 areas, yielding total feature counts of 2,000--20,000 across runs. 
% %
% In the second test, spatial join performance was evaluated using all 28 neighborhood polygons as the fixed reference layer, joined against a dataset of 311 service registers containing approximately 4.5 million georeferenced points, sampled at 10\% increments to assess\footnote{\url{https://shorturl.at/HylCQ}}.
% %
% Finally, in the third test, the GPU compute engine's performance was evaluated using the entire Manhattan road network to run render-based computations across approximately 11,000 street segments.
% %
% We run each test 10 times to assess the system's robustness.
%
Three performance experiments were conducted on an M1 Mac Studio (32GB RAM) to evaluate the feasibility of \autark's serverless approach.
The first assessed loading time by incrementally loading Manhattan neighborhood data from 1 to 10 areas (2,000--20,000 features). The second evaluated spatial join performance using 28 neighborhood polygons joined against approximately 4.5 million georeferenced 311 service records, sampled at 10\% increments\footnote{\url{https://shorturl.at/HylCQ}}. The third evaluated GPU compute performance across approximately 11,000 segments of Manhattan's road network.
Each test was run 10 times to assess robustness.

% Figure~\ref{fig:performance} reports the results. 
For loading time, at 20,000 features, the total time averages 6.0s (sd=460ms), with database ingestion accounting for 4.9s (sd=295ms) and 3D map data loading for 1.1s (sd=180ms), yielding a cost of $\approx$0.3ms per feature. 
For spatial joins, processing the full 4.5 million records requires, on average, 2.6s (sd=6ms), corresponding to $\approx$0.58$\mu$s per record. Both metrics scale linearly with dataset size, suggesting predictable behavior for larger inputs. 
%
% The sky view factor computation processed 11,224 road segments across Manhattan in approximately 5.3 seconds total. The CPU-side work was marginal: triangulating the building geometry took 74.5 ms, allocating GPU buffers and compiling pipelines took 36.6 ms, and encoding over 11,000 render commands took only 16.7 ms — together, less than 130 ms, or roughly 2\% of the total runtime. The remaining 5.18 seconds were spent waiting for the GPU to execute all render passes and for the result buffer to be mapped back to CPU memory. The computation allocated a single 6,784 × 6,784 RGBA8 render target to hold all 11,224 tiles simultaneously, consuming approximately 185 MB of GPU memory for the texture alone, in addition to the building vertex and index buffers. Each tile is 64 × 64 pixels, giving 4,096 pixels per viewpoint from which the sky view factor is derived as the fraction of pixels not covered by building geometry.
% 
% For the render-based computation, the test allocated a single $6,784 \times 6,784$ render target to hold 11,224 tiles ($64 \times 64$ pixels each), consuming 185 MB of GPU memory, and ran on average in 4.2s (sd=45ms).
Render-based computation test used a $6,784 \times 6,784$ target to store 11,224 tiles ($64 \times 64$ pixels), consuming 185 MB of GPU memory and ran on average in 4.2s (sd=45ms).

These results are particularly noteworthy given that \autark runs entirely in the browser, where such operations would traditionally require a backend, and that no optimizations such as data streaming were applied. Given that modern browsers can allocate several gigabytes of memory per tab, these results suggest that serverless architectures are a feasible alternative for urban visual analytics systems.
%
% \begin{figure}
%   \centering\includegraphics[width=\linewidth]{figs/performance_render.png}
%   \caption{\autark's performance benchmarks. (Top)~Loading time as a function of the number of processed OSM data features, broken down into database and 3D map ingestion contributions. (Bottom)~The time required to perform spatial join operations for a fixed set of polygons and an increasing number of data points.}
%   % \marcos{Compute + Spatial Join.}
%   \label{fig:performance}
% \end{figure}

\subsection{Agentic urban VA development with \autark}
The development of Urban VA systems is undergoing a paradigm shift in which code is increasingly written by agents, yet the resulting codebase must remain understandable and maintainable by humans who define requirements and verify correctness. This raises a key question: how does a toolkit's design influence the quality of agent-generated code?
To investigate this, we evaluated \autark's effectiveness through a controlled experiment comparing agent-based implementations of well-scoped urban VA systems across two conditions: one constrained to \autark's modules and one that allowed the agent to use any general-purpose libraries available online (baseline). Our goal was twofold: to assess whether agents could deliver functional applications in both conditions, and to evaluate the structural properties of the resulting code with respect to complexity, maintainability, and readability.

\myparagraph{Experimental setup} We designed five urban VA tasks of increasing complexity, each requiring a combination of data ingestion, spatial operations, and interactive visualization. The first task, the simplest, requires loading multiple OpenStreetMap Layers and coloring specific layers based on a spatially joined attribute from an external thematic dataset. 
Each successive task adds requirements: additional thematic datasets, multiple spatial joins, linked charts, and GPU-accelerated computations. All the tasks, prompts, and experiment results are available in the supplemental material of this paper on the GitHub\footnote{\url{https://github.com/urban-toolkit/autark-experiments}}. Each task was implemented by an AI coding agent (Claude Code with Opus 4.6 model). 
For both conditions, we prompted the task definition and used the same model configuration. Each trial was conducted independently, with no conversation history carried over between tasks.
To ensure a fair comparison, both conditions shared the same stop criteria: after generating the project, the agent was required to validate that the system fully compiles, builds, and serves without errors.
For the \autark condition, the agent was provided with \autark's documentation as its primary context and instructed to use only \autark's API. For the general condition, the agent was explicitly told not to use \autark to build the system. 

\begin{figure}
  \centering\includegraphics[width=1\linewidth]{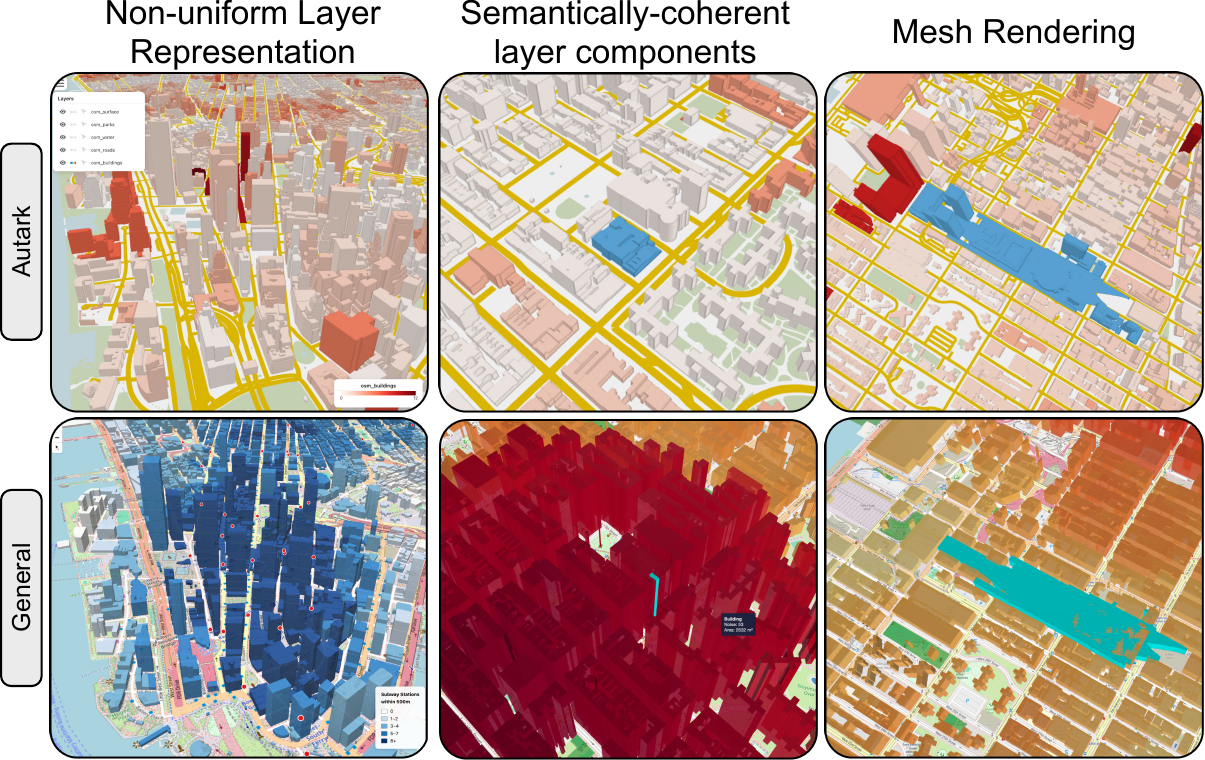}
  \caption{Comparison of agent-generated outputs between the \autark condition (top row) and the baseline (bottom row). (Left) The baseline combines multiple rendering libraries into a fragmented visual stack. (Center) The baseline treats parts of the same physical structure as independent geometries, whereas \autark reconstructs semantically coherent building footprints. (Right) The baseline produces artifacts, such as fragmented polygons, whereas \autark ensures artifact-free rendering.}
  \label{fig:llmresults}
\end{figure}

\myparagraph{Functional issues} Our first goal in analyzing the experiments was to assess whether the generated systems behave as specified. For each task and condition, we manually inspected the output application against its functional requirements, such as layer rendering, data encoding, spatial operations, and view coordination.
% Across the challenges imposed by each new task, we highlight deviations from the expected behavior. 
%
The first issue concerns layer representation. Without \autark, the agent frequently combines multiple mapping libraries within the same application to render different urban layers. As shown in the leftmost column of Figure~\ref{fig:llmresults}, this results in a fragmented visual stack, with layers from different rendering libraries coexisting on the same canvas. 
Beyond the visual inconsistency, it introduces a functional limitation: picking operations cannot traverse libraries, making it impossible to select or interact with features on layers that are occluded by elements rendered by a different backend. \autark avoids this entirely by routing all layers through a single rendering pipeline, ensuring that picking and interaction work uniformly across the full layer stack.
The second issue concerns layer processing. \autark's spatial database module processes raw OpenStreetMap data through a unified pipeline that reconstructs semantically coherent geometries, assembling building footprints from their constituent way segments before storing them as individual features. Examples from the baseline condition lack this consolidation step. As a result, parts of the same physical structure are treated as independent geometries, as illustrated in the center column of Figure~\ref{fig:llmresults}. This inconsistency hinders geometry-dependent computations, as thematic attributes might be assigned to parts of the layer elements, and physical properties, such as building areas, become unreliable. These processing issues also affect rendering quality. As shown in the rightmost column of Figure~\ref{fig:llmresults}, the baseline condition produces visible mesh artifacts, such as fragmented polygons and overlapping edges that result from operating on raw, unconsolidated geometry. \autark's unified processing pipeline reconstructs coherent geometries before rendering, ensuring that the visual output is free of such artifacts.

% \joao{Still need to add GPU computation made easier with Autark} \joao{Refer back to shadow computation when talking about the defective meshes.} \joao{Uma porrada de rua nada a ver}

\myparagraph{Code generation metrics} 
Beyond functional correctness, we also profile the generated code to characterize its structural properties. Even in an agentic workflow, developers must be able to read, navigate, and understand the generated codebase in order to verify the agent's design decisions. To this end, we computed multiple code metrics that serve as proxies potentially indicative of code understandability and ease of maintenance. Per-trial results are available in the supplemental material. \autark-based implementations average 153~LOC compared to 590 for the general condition and are consistently contained in a single source file, whereas the general condition spreads across 6~files on average. The general condition also introduces roughly twice as many external dependencies (6.1 vs.\ 3.1 on average, \autark-based code mainly imports its own modules), each of which adds API surface that the developer must understand, version constraints that must be managed, and potential points of failure that complicate debugging, all of which compound the maintenance burden of the generated codebase. Cyclomatic complexity measures the number of independent paths through a program's control flow, serving as a widely used proxy for how difficult code is to understand and test~\cite{mccabe1976complexity}. Higher values indicate more branching logic that developers must trace when verifying correctness. \autark code averages a cyclomatic complexity of 22 compared to 84 for the general condition, indicating that the toolkit's declarative abstractions absorb much of the branching logic that the agent must otherwise generate from low-level primitives.
Taken together, these metrics suggest that Autark's domain-specific abstractions not only reduce the volume of generated code but also produce codebases that are structurally simpler and more uniform. As demonstrated in the usage scenarios (Section~\ref{subsec:usage_examples}), these gains do not come at the cost of expressiveness, as \autark achieves feature-parity with traditional urban VA systems while producing code that is more compact. While these proxies do not guarantee maintainability, they consistently lower the barrier for developers who need to inspect, verify, and extend the agent-generated codebase.

%% file: tex/08-conclusion.tex
\begin{figure}[H]
  \centering\includegraphics[width=\linewidth]{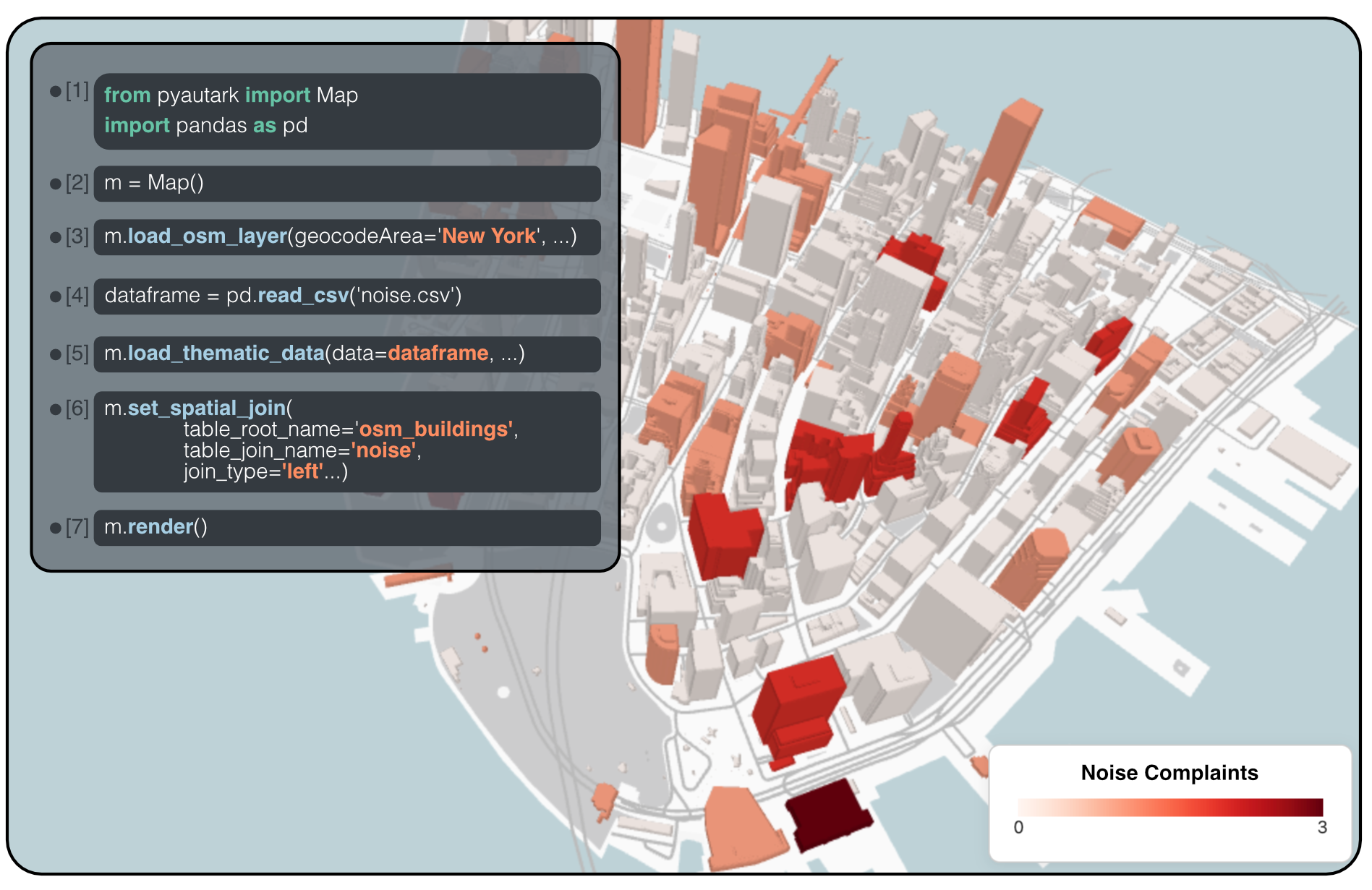}
  \caption{\autark wrapper for Jupyter notebooks, illustrating how its serverless architecture facilitates integration with diverse computational environments and community standards.}
  \label{fig:pyautark}
\end{figure}

\section{Conclusion}
\myparagraph{Community adoption} \autark's self-contained, serverless architecture facilitates integration with diverse computational ecosystems without requiring specific development environments or deployment stacks. Figure~\ref{fig:pyautark} shows \autark used within a Jupyter notebook via a Python wrapper (pyautark\footnote{\url{https://github.com/urban-toolkit/pyautark}}), analyzing buildings affected by noise complaints in New York's Financial District. Beyond Jupyter, \autark is compatible with cloud-hosted platforms such as Google Colab, reactive notebook environments such as Observable~\cite{observablehq}, and WebXR platforms for immersive analytics~\cite{wagner2024reimagining}.

\myparagraph{Limitations} \autark's serverless architecture, while effective for rapid prototyping, introduces inherent scalability constraints. The toolkit is bound by the memory and compute resources available in a single browser tab. For the datasets explored in this paper, this proved sufficient, with linear scaling behavior and interactive response times (Section~\ref{subsec:performance}). However, for terabyte-scale datasets, a browser-based approach is not a viable alternative to server-backed architectures with dedicated database infrastructure. Additionally, the set of abstract chart types currently provided by \autark was informed by a decade of experience building urban VA systems and by a thorough review of the urban VA literature, but it is not exhaustive. \autark partially addresses this by providing an extensibility mechanism that allows developers to define custom visualization code while retaining integration with the feature-centric model. A more comprehensive built-in chart library remains a direction for future work.

\myparagraph{Future work} We envision two main directions for future work. First, we plan to leverage \autark to better understand the construction of VA systems within agentic ecosystems. The evaluation presented in this paper adopted a purely agentic approach, where the agent was solely responsible for generating the entire codebase. In practice, however, the development process is more nuanced: developers may co-write code alongside agents, selectively delegate specific components, or write entire modules themselves. We plan to conduct a large-scale user study using \autark to investigate how developers interact with agents across this spectrum of collaboration, examining how toolkit design influences not only the quality of the output but also the dynamics of the human-agent development process. Second, we plan to explore integrating \autark into immersive environments. \autark's web-first, serverless architecture is inherently compatible with WebXR, and its feature-centric model provides a natural foundation for immersive urban analytics. This direction has the potential to reduce barriers to developing immersive VA applications, which currently require substantial additional engineering effort beyond an already complex urban VA stack.